\documentclass{ws-ijmpe}

\def\etal{{\it et al.}}

\newcommand{\br}{\hbox{\bf r}}
\newcommand{\rhoi}{\rho^{i/3}}
\newcommand{\Ps}{P_{\sigma}}

\newcommand{\RR}{\hbox{\bf R}}

\begin{document}

\markboth{B.~Cochet, K.~Bennaceur, J.~Meyer, P.~Bonche
and T.~Duguet}{Skyrme forces with extended density dependence}

%
\catchline{}{}{}{}{}
%

\title{SKYRME FORCES WITH EXTENDED DENSITY DEPENDENCE
}

\author{\footnotesize B.~COCHET,
K.~BENNACEUR\footnote{bennaceur@ipnl.in2p3.fr},
J.~MEYER
}

\address{IPN Lyon, CNRS-IN2P3/UCB Lyon 1\\
43, Bd. du 11 novembre 1918\\
69622 Villeurbanne cedex, France\\}

\author{P.~BONCHE, T.~DUGUET
\footnote{present address: Physics Division, Argonne National Laboratory,
Argonne, IL 60439, USA}}

\address{Service de Physique Th\'eorique, CEA Saclay\\
91191 Gif-sur-Yvette cedex, France}

\maketitle

\begin{history}
\received{(received date)}
\revised{(revised date)}
\end{history}

\begin{abstract}
A generalized parameterization of the Skyrme effective force is discussed. 
Preliminary results are presented for infinite symmetric and asymmetric 
nuclear matter. 
In particular, it is shown that an enlarged density dependence based on 
two terms allows to choose independently the incompressibility and the 
isoscalar effective mass.
\end{abstract}

%
%

\section{Introduction}

It is commonly accepted that it does exist a relation between
compressibility, effective mass and the density dependence of a given
effective force.
Studies with Skyrme forces~\cite{sly4} have shown that the incompressibility
$K_\infty$ and the effective mass cannot be chosen independently
once the analytical form of the (single) density-dependent term has been
chosen.
This has led to the $\rho^{1/6}$ density dependence in Skyrme forces
like SkM$^*$ which allows a value of $K_\infty$ around $220$~MeV
close to that extracted from the experimental breathing mode
analyses~\cite{blaizot_rep,blaizot_95} and an effective mass $m^*$ around
$0.7\,m$ simultaneously. 
A general review of the knowledge about
compressibility has been recently summarized by
G.~Col\`o {\it et al.}~\cite{cologiai}.

Recently, the density dependence of phenomenological effective interactions,
such as Skyrme or Gogny forces, has been revisited in the context of beyond
mean field calculations~\cite{duguet1,duguet2}.
Indeed, while a dependence of the interaction on the density
is well established for calculations at the mean field level~\cite{NV72},
no strongly motivated prescription exists when several mean fields are
mixed as in the Generator Coordinate Method (GCM) and the
Projected Mean Field Method (PMFM). First, an extension of the 
Goldstone-Brueckner theory has motivated
the GCM and the PMFM from a perturbative point of view for the first
time~\cite{duguet1}.
In this extended context, a generalized Brueckner $G$ matrix summing
particle-particle ladders has been defined and may be used as a reference
from which phenomenological interactions in GCM or PMFM calculations
should be approximated.
It is possible to simplify this in-medium interaction to
extend the validity of the Skyrme force~\cite{skyrme,VB72} and to
identify approximately the density dependence originating from Brueckner
correlations in the context of mixed nonorthogonal vacua~\cite{duguet2}.
Consequently, a single density-dependent term has often been used in
phenomenological forces, e.g. Gogny~\cite{decharge} and Skyrme~\cite{VB72}
effective interactions.
On the other hand, the density dependence re-normalizing three-body force
effects has
been shown to be different from the one taking care of Brueckner correlations
when going beyond the mean field approximation~\cite{duguet2}.
Based on this analysis, it is legitimate to redefine the Skyrme
interaction at the mean field level.
Including two density-dependent terms, each related to a given physical
origin,  will allow a proper extrapolation of the interaction to GCM and
PMFM calculations.

A first attempt is presented in this article, where
we concentrate on the practical
advantages of having two different density-dependent terms.
Our analysis is focused on $\rho^{1/3}$ terms which are connected with
a $k_F$ expansion of the Brueckner $G$ matrix.

%
%

\section{Standard Skyrme effective interactions and beyond}

%
\begin{figure}[th]
\centerline{\psfig{file=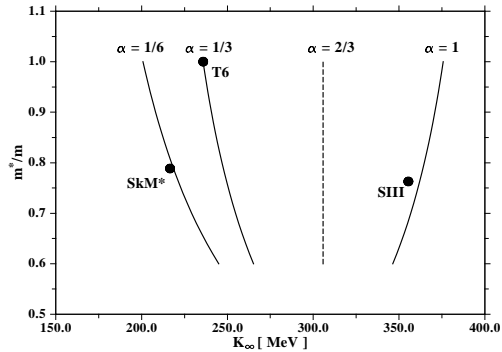,width=4.6cm,angle=270}}
\vspace*{8pt}
\caption{Correlation between incompressibility, effective mass and
density dependence of standard parameterizations of Skyrme effective
interactions.
When $\alpha=2/3$, the value of $K_\infty$ is fixed and the effective 
mass can be freely chosen.
However, this case is not interesting since it does not correspond to a
realistic value of $K_\infty$.}
\label{fig:corrKm}
\end{figure}
%

When using Skyrme effective forces with a standard density dependence
$\rho^\alpha$, where $\rho$ is the total density, the adjustable
parameter $\alpha$ is strongly correlated to the
incompressibility and the effective mass in nuclear matter.
This is illustrated on Figure~\ref{fig:corrKm} for different standard
Skyrme forces with various powers $\alpha$.
In all cases, the different parameters of the force have been adjusted
in order to obtain the saturation density $\rho_0=0.16$~fm$^{-3}$ and
the energy per particle $E/A=-16$~MeV in infinite symmetric nuclear matter.
Once these two properties are fixed, the relation
between $m^*/m$ and $K_\infty$ is entirely determined by $\alpha$.
This well-known feature, discussed in ref.~\cite{sly4} yields to the
conclusion that a value of $\alpha$ around 1, like in SIII, does not 
allow to reach a correct incompressibility~\cite{blaizot_rep,giant}.
Only values of $\alpha$ ranging from $1/6$ to $1/3$ allow for an
acceptable set $\{m^*/m,K_\infty\}$.

This constraint disappears 
with the family of parameterizations presented in~\ref{sec:app}.
Following the previously discussed procedure to adjust the coefficients,
but having the additional flexibility coming from the second density
dependent term, we can fix
the saturation properties of infinite symmetric nuclear matter
$\rho_0$, $E/A$, $K_\infty$ and $m^*/m$ independently and determine the
parameters of the force.

However this procedure only guarantees to have realistic
properties of nuclear matter in the vicinity of the saturation
point. If we want a force which gives an equation of state
in agreement with {\it ab initio} calculations at low
density as well as high density (up~to~$\sim\,3\rho_0$)
one should adopt a different fitting procedure. First, one can fit
a set of points \smash{$\left[\rho_i,E/A(\rho_i)\right]_{i=1,...,N}$}
which samples a realistic equation of state of nuclear matter.
In this kind of approach, the presence of a density-dependent
term with the power $2/3$ in the force is particularly
interesting. Indeed, the
effective mass is not constrained by the fit and thus, can be
freely chosen~\cite{cbbdm}.

%
%


%
\begin{figure}[th]
\centerline{%
\psfig{file=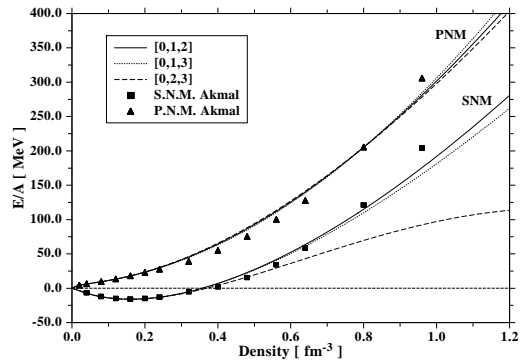,width=4.7cm,angle=270}
}
\vspace*{8pt}
\caption{Energy per particle in infinite matter as a function
of equilibrium density.
SNM: symmetric nuclear matter; PNM: pure neutron matter. Labels
[0,1,2], [0,1,3], [0,2,3] refers to the density dependence
of the forces (see text).
Full squares and triangles: EOS of Akmal {\it et al.}~$^{13}$.}
\label{fig:eos}
\end{figure}
%

The equations of state (EOS) obtained
using the generalized
forces $[0,1,2]$, $[0,1,3]$ and $[0,2,3]$ (see~\ref{sec:app})
for symmetric nuclear matter as well as pure neutron matter
are presented here.
The quantities $\rho_0=0.16$~fm$^{-3}$, $E/A=-16$~MeV, 
$K_\infty=230$~MeV, $m^*/m=0.8$ and $\kappa=0.5$
along with a fit of the neutron
matter equation of state provided by Akmal {\it et al.}~\cite{akmal}
allow the determination
of the parameters of the force.
Figure~\ref{fig:eos} shows
the two EOS (energy per particle as a function of equilibrium density)
for symmetric nuclear matter and pure neutron matter
compared with the EOS of Akmal \etal~\cite{akmal}.
The results are quite reasonable and rather similar for
densities from 0 to  $\sim 3\rho_0$.

%
%

\section{Conclusion}

We have shown that the choice of a standard Skyrme effective force with
a modified density dependence based on two terms enables to choose
independently the incompressibility and the isoscalar effective mass.
The gross properties of nuclear matter investigated here are
quite reasonable.
With this generalized density dependence we have been able to construct
new Skyrme like parameterizations without collapse at relevant densities
and which exhibit
reasonable equation of state (EOS) for symmetric nuclear
matter as well as for pure neutron matter compared to the recent
realistic variational EOS of Akmal~\etal~\cite{akmal}.

Other choices can be explored for the density dependence provided they
are physically well grounded.
The choice $\rho^{2/3}$ for one of the density-dependent term lets
total freedom for the effective mass. This feature is extremely
interesting from the perspective of developing accurate and
predictive force since it gives a control on the density of state
around the Fermi energy where the correlations beyond the Hartree-Fock
approximation can develop. Furthermore it implies only two
additional parameters so that their total number remains
quite small.

%
%

\appendix

\section{Extended parameterizations of Skyrme forces}
\label{sec:app}

The parametrization of the Skyrme forces discussed in this
article is
\begin{equation}
V(\br_1,\br_2) =
          \sum_i \, t_{0i} \, \rhoi(\RR) \, (1+x_{0i}\Ps) \, \delta(\br)
          \ + \ \mbox{standard non local terms}
\label{eq:sky_full}
\end{equation}
%
%
In this first exploratory
work, we limit our study to the case of density dependent local terms
and three
non zero values for $t_{0i}$, namely $i_1$, $i_2$ and $i_3$ out of the set
$\{0,1,2,3\}$. The corresponding forces will be labeled by
$[i_1,i_2,i_3]$.

%


\end{document}